\documentclass[12pt]{iopart}

\usepackage{color}
\usepackage{amssymb}
\usepackage{graphicx}

\begin{document}

\title[Nonlinear hydrodynamic theory of crystallization]{Nonlinear hydrodynamic theory of crystallization}

\author{Gyula I. T\'oth}
\address{Institute for Solid State Physics and Optics, Wigner Research Centre for Physics,\\P.O. Box 49, H-1525 Budapest, Hungary}
\ead{toth.gyula@wigner.mta.hu}
\author{L\'aszl\'o Gr\'an\'asy}
\address{Institute for Solid State Physics and Optics, Wigner Research Centre for Physics,\\P.O. Box 49, H-1525 Budapest, Hungary}
\address{BCAST, Brunel University, Uxbridge, Middlesex, UB8 3PH, United Kingdom}
\author{Gy\"orgy Tegze}
\address{Institute for Solid State Physics and Optics, Wigner Research Centre for Physics,\\P.O. Box 49, H-1525 Budapest, Hungary}

\begin{abstract}
We present an isothermal  fluctuating nonlinear hydrodynamic theory of crystallization in molecular liquids. A dynamic coarse-graining technique is used to derive the velocity field, a phenomenology, which allows a direct coupling between the free energy functional of the classical Density Functional Theory and the Navier-Stokes equation. Contrary to the Ginzburg-Landau type amplitude theories, the dynamic response to elastic deformations is described by parameter-free kinetic equations. Employing our approach to the free energy functional of the Phase-Field Crystal model, we recover the classical spectrum for the phonons and the steady-state growth fronts. The capillary wave spectrum of the equilibrium crystal-liquid interface is in a good qualitative agreement with the molecular dynamics simulations.
\end{abstract}


\maketitle

\section{Introduction}

A general molecular scale theory of crystallization that describes elastic and plastic deformations, including  phonons, translation/rotation of the crystallites, and crack propagation is yet unavailable. Practical problems, like dendrite fragmentation or a realistic model of defect dynamics in crystals would require such a tool. Despite recent steps made in this direction, further work is needed. The advanced models developed in the past years are based on various formulations of the classical density functional theory (DFT) \cite{E,O,Si}, such as the perturbative approach by Ramakrishnan and Yussouff \cite{RY}, the non-perturbative theories like the weighted and modified weighted density approximations \cite{WDA,MWDA}, and ultimately the fundamental measure theory \cite{FMT}, followed by the recently proposed simple phase-field crystal (PFC) model \cite{PFC0,PFC1,AP}. The DFT models characterize the local state of the matter by the time averaged one-particle density of which the free energy is a unique functional. The dynamic extensions of the DFT, termed dynamic density functional theories (DDFTs), address a great variety of crystallization related time dependent problems \cite{AP}, including grain boundary dynamics, homogeneous and heterogeneous nucleation, dendritic and eutectic growth, density- or solute trapping, freezing of competing crystalline phases, etc. \cite{AP}. 

The dynamic extension of the DFT even for simple atomic liquids is a rather demanding problem of non-equilibrium statistical physics, posing a number of difficulties \cite{HGL,MT}. First the DDFT was worked out for strongly non-equilibrium Brownian liquids and colloidal systems \cite{MT,AE,AR,TLL}. Later the dynamics was extended for dense atomic liquids \cite{A06}. The unified dynamics derived for both molecular and colloidal fluids \cite{A09} is a recent development, but it is valid only for small velocities (i.e. for dense liquids and/or close to equilibrium). The first time derivative in the dynamical equation originates from the presence of a medium, which is responsible for the damping. In the case of simple molecular liquids no medium is present, thus the velocity field cannot be neglected anymore. Some of the dynamic extensions to crystals rely on a purely diffusive dynamics (overdamped, conservative kinetic equation yielding diffusive dynamics, more appropriate for colloidal suspensions than anything else, e.g., \cite{PFC0,PFC1,AP,TBVL}), whereas other models incorporate a second order time derivative in the kinetic equation \cite{SHP}, which may lead to the appearance of a second time scale resembling  relaxation. A third category is the linearized hydrodynamic model that would capture  phonons in the crystal \cite{MG,MNG}. However, only a few limited studies of the latter kind have been performed \cite{MNG}, in which the form of the associated phonon dispersion has not been tested. 

A few other attempts have recently been made to couple fluid flow and the PFC model. Pretorius {\it et al.} \cite{PV} proposed a simple advected PFC model to describe particle motion in a carrier fluid. A Navier-Stokes-Cahn-Hilliard model has been coupled with the PFC model to model jamming of phase separation by colloidal particles at the liquid-liquid interface \cite{ALV11,ALV12}. This approach, however, also lacks the fluid flow for the density field.

The most essential problem of combining a molecular scale theory with hydrodynamics is that in the latter approach inertia and dissipation are formulated on the macroscopic level, which cannot be easily reconciled with molecular scale modeling (e.g., the large density gradients in the microscopic picture should not create spurious mass flow inside the crystal). In this paper, we raise the idea that a straightforward resolution of this problem could be obtained via scale separation; i.e., by employing appropriately coarse-grained quantities in the Navier-Stokes equation. 

Herein, we present a nonlinear hydrodynamic theory of crystallization in simple molecular liquids that naturally incorporates elasticity and plasticity including  phonon dispersion, capillary waves, together with translation and rotation of the crystal embedded into the liquid. This is achieved by coupling the Navier-Stokes equation with classical density functional theories of crystallization via the coarse-grained density and velocity fields. Our paper is structured as follows: In section 2, we present the dynamical equations and the coupling between microscopic and macroscopic quantities on the basis of scale separation. To be able to address fluctuation phenomena (such as capillary waves at the crystal-liquid interface), we start from Fluctuating Nonlinear Hydrodynamics (FNH), a field-theoretic approach that couples hydrodynamic fluctuations with the free energy \cite{SVC}. We then demonstrate the translational and rotational invariance of our model. In section 3, we apply our approach to a simple classical DFT, the PFC model, and specify the respective model parameters for pure iron. Section 4 deals with the dynamic response to elastic deformations, including compression, torsion and  phonon spectrum, followed by results for front propagation for growth and melting, and the capillary wave spectrum. Finally, a few technical remarks are made regarding coarse-graining applied in this work. In section 5, we offer a few concluding remarks.

\section{Model description}

\subsection{Dynamical equations}

\par We present a time-dependent Density Functional Theory of crystallization based on Fluctuating Nonlinear Hydrodynamics \cite{LL,NY}. Our starting point is a recent work of Shang {\it et al.} \cite{SVC}, which has successfully addressed multiphase liquid and liquid-gas equilibrium interfaces. Considering isothermal processes, the respective dynamical equations read as:
\begin{eqnarray}
\label{eq:eq3}\frac{\partial \rho}{\partial t} + \nabla \cdot \mathbf{p}  &=& 0 \\
\label{eq:eq4} \frac{\partial \mathbf{p}}{\partial t} + \nabla \cdot \left( \mathbf{p} \otimes \mathbf{v} \right) &=& \nabla \cdot [ \mathbb{R}(\rho) + \mathbb{D}(\mathbf{v}) + \mathbb {S} ] \enskip .
\end{eqnarray}
Here $\rho$ is the mass and $\mathbf{p}$ the momentum density, and $\mathbf{v}$ the velocity field. $\mathbb{R}(\rho)$ and $\mathbb{D}(\mathbf{v})$ are the reversible and dissipative stress tensors, respectively, whereas $\mathbb{S}$ is a noise matrix representing thermal fluctuations. The dissipative stress of a Newtonian liquid is as follows:
\begin{equation}
\label{eq:eq6}
\mathbb{D}(\mathbf{v}) = \mu_S [ (\nabla \otimes \mathbf{v}) + (\nabla \otimes \mathbf{v})^T ] + \left( \mu_B - \frac{2}{3} \mu_S \right) (\nabla \cdot \mathbf{v}) ] \enskip ,
\end{equation}
where $\mu_S$ and $\mu_B$ are the shear and bulk viscosities, respectively. The fluctuation-dissipation theorem yields the following covariance tensor for the noise:
\begin{eqnarray}
\label{eq:eq7}
\langle S^{\mathbf{r},t}_{ij}S^{\mathbf{r}',t'}_{kl}\rangle &=& \left( 2 k_B T \mu_S \right) \delta(\mathbf{r}-\mathbf{r}') \delta(t-t') \times \\
\nonumber && \times \left[ \delta_{ik}\delta_{jl}+\delta_{jk}\delta_{il} + \left( \frac{\mu_B}{\mu_S} - \frac{2}{3} \right) \delta_{ij}\delta_{kl} \right] \enskip ,
\end{eqnarray}
where $k_B$ is Boltzmann's constant and $T$ the temperature. The free energy functional of the classical DFT is anchored to the FNH model via the fundamental equation for the reversible stress tensor emerging from the \textit{least action principle} \cite{Sa}:
\begin{equation}
\label{eq:eq5}
\nabla \cdot \mathbb{R}( \rho) = - \rho \nabla \frac{\delta F[\rho]}{\delta \rho} \enskip ,
\end{equation}
where $\delta F[\rho]/\delta \rho$ is the first functional derivative of the Helmholtz free energy functional $F[\rho]$ with respect to the density.\\
Note that substituting the time derivative of equation (\ref{eq:eq3}) into the divergence of equation (\ref{eq:eq4}), and using equation (\ref{eq:eq5}) yields the following kinetic equation:
\begin{equation}
\label{eq:arch}
\frac{\partial^2 \rho}{\partial t^2} = \nabla \cdot \left( \rho \nabla \frac{\delta F}{\delta \rho }\right) + \nabla \cdot [\mathbb{D}(\mathbf{v})-(\nabla \otimes \mathbf{v})+\mathbb{S}] \enskip ,
\end{equation}
which consistently recovers the \textit{undamped} limit of the dynamics proposed by Archer \cite{A09} for small velocities ($\mathbf{v}\to 0$) in the case of $\mathbb{S}=0$. Since in equation (\ref{eq:arch}) $\rho$ and $\mathbf{p}$ are the local time-averaged microscopic one-particle and momentum densities, respectively, the same applies for equations (\ref{eq:eq3}) and (\ref{eq:eq4}).

\subsection{Scale separation via dynamic coarse-graining}

\par The main difficulties associated with applying the free energy functional of a classical DFT of freezing \textit{directly} in the dynamical equations (\ref{eq:eq3}) and (\ref{eq:eq4})  are as follows: (i) The validity limit of the Navier-Stokes equation coincides with that of the linear dissipation assumption. According to molecular dynamics simulations, the smallest length scale, where the Navier-Stokes equation is considered to be valid in liquids is $\approx 10^2 \times \sigma_0$, where $\sigma_0$ is the effective molecular diameter \cite{LiLi}. On smaller length scales, the definition of the continuum velocity field breaks down, and the system shows an essentially different behavior (in other words, the molecular resolution must be taken into account in both space and time). (ii) In the classical DFT of freezing, the crystal is represented by a lattice periodic density distribution, which shows peaks at the lattice sites. Due to the high gradients of the density field, expressing $\mathbf{v}$ by $\mathbf{p}$ and $\rho$ may lead to singularities in the velocity, yielding spurious convections in the interatomic space. To prevent such singularities while retaining the validity of the Navier-Stokes equation, we assume that the convection is related to the coarse-grained momentum and density fields as follows:
\begin{equation}
\label{eq:defvel}
\mathbf{v}(\mathbf{r},t) := \hat{\mathbf{p}}(\mathbf{r},t)/\hat{\mathbf{\rho}}(\mathbf{r},t) \enskip ,
\end{equation}
where $\hat{\rho}(\mathbf{r},t)$ and $\hat{\mathbf{p}}(\mathbf{r},t)$ stand for the space- and time-averaged  microscopic one-particle density and momentum fields, respectively, where the spatial averaging results in the fields \textit{without their lattice periodic components}. Consequently, this phenomenological "derivation" relies on a two-step coarse-graining process. Note that the shortest possible wavelength in $\mathbf{v}(\mathbf{r},t)$, on which the Navier-Stokes equation is valid, corresponds to the thickness of the solid-liquid interface that should be several times the intermolecular distance, a condition not necessarily satisfied for faceted crystal-liquid interfaces. In addition, we apply the constant mobility approximation \cite{TBVL} in equation (\ref{eq:eq5}):
\begin{equation}
\label{eq:pressure}
\nabla \cdot \mathbb{R(\rho)} = - {\rho_0}\nabla \frac{\delta F[\rho]}{\delta \rho} \enskip ,
\end{equation}
which corresponds to the trivial reversible stress tensor $\mathbb{R}(\rho)=-\rho_0(\delta F[\rho]/\delta\rho)\mathbb{I}$ (where $\mathbb{I}$ is the identity matrix). The coarse-grained velocity can be generated by using different lowpass filters. The properties and the effect different filters will be discussed in section 4.5.

Note that applying our approach to a simple Ginzburg-Landau (GL) type free energy functional (where the preferred density field distributions are homogeneous), the coarse-graining process simply results in $\hat{\rho}\approx\rho$, $\hat{\mathbf{p}}\approx\mathbf{p}$ and $\mathbf{v}\approx\mathbf{p}/\rho$ [if the characteristic length of the GL model (i.e. the interface width) is large compared to the interatomic distance]. In this case, the model reduces to the description of Shang et al. \cite{SVC}, but with constant mobility in the reversible stress. (Since the GL models do not contain the elastic properties of the crystalline phase, the reduced model can be used only for liquid-liquid or liquid-vapor transitions.) Finally, we also mention that the present description is limited exclusively by the ratio of the interatomic distance and the interface thickness: the characteristic length scale on which the viscous dissipation acts must be large enough to retain the validity of the Navier-Stokes equation. Molecular dynamics investigations along the line described in Ref. \cite{LiLi} are needed to clarify this limitation further.

\subsection{Translational and rotational invariance}

First we test our description against translational and rotational invariance. Pure translation and/or rotation can be expressed by the velocity field $\mathbf{v}_0(\mathbf{r}) = \mathbf{v}_0 + \mathbf{\omega} \times (\mathbf{r}-\mathbf{r}_0)$, while the corresponding momentum distribution reads as $\mathbf{p}(\mathbf{r,t})=\rho(\mathbf{r},t) \cdot \mathbf{v}_0(\mathbf{r})$. Using equation (\ref{eq:defvel}) the coarse-grained velocity field is:
\begin{equation}
\mathbf{v}(\mathbf{r},t)=\frac{\hat{\mathbf{p}}(\mathbf{r},t)}{\hat{\rho}(\mathbf{r},t)}=\frac{\hat{\rho}(\mathbf{r},t) \cdot \mathbf{v}_0(\mathbf{r})}{\hat{\rho}(\mathbf{r,t})}\equiv \mathbf{v}_0(\mathbf{r}) \enskip ,
\end{equation}
where we utilized that the spectrum of $\mathbf{v}_0(\mathbf{r})$ contains only a zero-frequency component.

Consider now an equilibrium density distribution $\rho_0(\mathbf{r})$, i.e. for which the first functional  derivative of the free energy functional is constant: $\{\delta F[\rho(\mathbf{r})]/\delta \rho\} |_{\rho_0(\mathbf{r})}=\mu$. The functional derivative shows translational and rotational invariance, namely: $\{\delta F[\rho(\mathbf{r})]/\delta \rho\} |_{\rho(\mathbf{r},t)}=\mu$ for $\rho(\mathbf{r},t)=\rho_0[\mathbf{r}-\mathbf{v}_0(\mathbf{r}) \cdot t]$. Using this together with $\mathbf{p}(\mathbf{r,t})=\rho(\mathbf{r},t) \cdot \mathbf{v}_0(\mathbf{r})$ and $\mathbf{v}(\mathbf{r})=\mathbf{v}_0(\mathbf{r})$ in equations (\ref{eq:eq3}) and (\ref{eq:eq4}) yields:
\begin{eqnarray}
\label{eq:simp1}\frac{\partial \rho(\mathbf{r},t)}{\partial t} + \nabla \cdot [\rho(\mathbf{r},t) \cdot \mathbf{v}_0(\mathbf{r})]  &=& 0 \\
\label{eq:simp2}\frac{\partial \mathbf{p(\mathbf{r},t)}}{\partial t} + \nabla \cdot \left[ \mathbf{p}(\mathbf{r},t) \otimes \mathbf{v}_0(\mathbf{r}) \right] &=& 0 \enskip ,
\end{eqnarray}
where we used that $\nabla\mu\equiv 0$ and $\mathbb{D}\{\mathbf{v}_0[\mathbf(r)]\}\equiv 0$. Note that equations (\ref{eq:simp1}) and (\ref{eq:simp2}) simply translate and rotate the initial density [$\rho_0(\mathbf{r})$] and momentum [$\mathbf{p}_0(\mathbf{r})=\rho_0(\mathbf{r})\cdot\mathbf{v}_0(\mathbf{r})$] distributions.

\subsection{Advantages of the model}

The novel scale separation method poposed here leads to the proper momentum equation in the long wavelength regime, while realizing wave dynamics for small wavelengths (the latter is responsible for the kinetics of the periodic field, i.e., freezing and melting). The frequency regimes are coupled via the nonlinear terms of the Navier-Stokes equation. The main advantages of the proposed dynamics to former descriptions are as follows: (i) As a direct  consequence of the scale separation, the present formalism is independent from the particular choice of the free energy functional, and (ii) the elastic properties of the crystal are inherently incorporated into the classical DFT free energy functional, while  the macroscopic liquid viscosity is coupled only to the coarse-grained velocity field. In contrast, in case of hydrodynamics-coupled Ginzburg-Landau type or phenomenological phase-field models the elastic crystal behavior is simply replaced by rigid body motion, which is usually achieved by applying a phase state dependent viscosity in the dissipative term. Unfortunately, this approximation violates both the linear dissipation approximation and the fluctuation-dissipation theorem at the interface.

\section{Application to the phase-field crystal model}

\subsection{Free energy functional}

\par We have chosen the free energy functional of a simple classical DFT, the Phase-Field Crystal theory (PFC) \cite{PFC0} to demonstrate the capabilities of our approach. Variants of this PFC model have been successfully used to describe elastic interactions in the crystal \cite{SHP,EG}, planar crystal front growth \cite{TGTPJAP}, faceting and branching during crystallization in colloidal systems \cite{TTG}, homogeneous \cite{TPTTG} and heterogeneous nucleation \cite{TTPG}, multiple time scales during glass formation \cite{BG}, or even solute trapping \cite{HHP}. (For a review, see \cite{AP}.) The free energy functional we use reads as \cite{JAEA}:
\begin{equation}
\label{eq:eq11}
\frac{F}{n_0 k_B T} = \int dV \left\{ f(n) - \frac{C_2}{2}(\nabla n)^2 - \frac{C_4}{2}(\nabla^2 n)^2 \right\} \enskip ,
\end{equation}
where the local free energy density is
\begin{equation}
\label{eq:eq12}
f(n)=(1-C_0)\frac{n^2}{2} - \left(\frac{a}{2}\right)\frac{n^3}{3} + \left(\frac{b}{3} \right)\frac{n^4}{4} \enskip .
\end{equation}
Here the order parameter is the normalized mass density: $n=\tilde{\rho}-1$, where $\tilde{\rho}=\rho/\rho_0$ ($\rho_0=m_0 n_0$ is the reference density, where $m_0$ is the atomic mass and $n_0$ the number density of the reference liquid). The parameter  $C_0$ can be related to the bulk modulus of the reference liquid via $K_0=(1-C_0)(n_0 k_B T)$, while $C_2$ and $C_4$ are responsible for elasticity. Using $n(\mathbf{r},t):=\bar{n}+\delta n(\mathbf{r},t)$ [where $\bar{n}$ is the average and $\delta n(\mathbf{r},t) \ll 1$] in equations (\ref{eq:eq11}) and (\ref{eq:eq12}) together with equation (\ref{eq:eq5}) yields the isothermal speed of sound $c$ in the liquid as the function of the average density: $c^2=(1-C_0- a \bar{n}+b \bar{n}^2)(k_B T/m_0)$, which recovers that of the reference liquid ($c_0=\sqrt{K_0/\rho_0}$) for $\bar{n}=0$. From the viewpoint of the numerical simulations it is convenient to scale the length by $\lambda=\sqrt{2 |C_4|/C_2}$, the free energy by $k_B T$, the density by $\rho_0$, and the time by $\tau=\sqrt{\rho_0 \lambda^5/(k_B T)}$ in the dynamical equations.

\subsection{Model parameters}

\par The physical parameters used in the simulations refer to a metallic system (a rough approximant of iron, see Table 1). With the present choice of the parameters, the PFC model prefers a homogeneous liquid - triangular crystal equilibrium in 2D  with the following equilibrium properties: the scaled average liquid density is $\tilde{\rho}_L^{eq}=1.051822$, the relative equilibrium density gap is $2(\tilde{\rho}_S^{eq}-\tilde{\rho}_L^{eq})/(\tilde{\rho}_S^{eq}+\tilde{\rho}_L^{eq}) = 0.1875\%$, where $\tilde{\rho}_S^{eq}$ is the equilibrium crystal density, whereas the scaled lattice constant is $\sigma_0 \approx 1.001\times (4\pi/\sqrt{3})$.

\begin{table}
\caption{Model parameters taken from Refs. \cite{JAEA}, \cite{Ka} and \cite{SVC}, denoted by [\textbf{a}], [\textbf{b}] and [\textbf{c}], respectively.}
\begin{tabular}{|c|c|c|c|c|}
\hline
$C_0$ & $C_2$ [\AA$^2$] & $C_4$ [\AA$^4$] & $a$ & $b$ \\
\hline
 $-10.9153^{[\mathbf{a}]}$ & $2.6^{[\mathbf{a}]}$ & $-0.1459^{[\mathbf{a}]}$ & $0.6917^{[\mathbf{a}]}$ & $0.0854^{[\mathbf{a}]}$ \\ 
\hline
\hline
 $T$ [K] & $n_0$ [1/\AA$^{3}$] & $m_0$ [a.m.u.] & $\mu_S^0$ [mPas] & ${\mu_B^0/\mu_S^0}$ \\
\hline
 $1833^{[\mathbf{a}]}$ & $0.0801^{[\mathbf{a}]}$ & 55.845 & $5.0^{[\mathbf{b}]}$ & $2^{[\mathbf{c}]}$ \\ 
\hline
\end{tabular}
\end{table}

\section{Results and discussion}

\par We performed illustrative simulations in two dimensions (2D) for (i) the elastic response to compression and torsion on an equilibrium crystalline cluster, (ii) planar crystal growth and melting fronts (without noise), and (iii) the capillary wave spectrum of the planar equilibrium crystal-liquid interface. The kinetic equations were solved using a pseudo-spectral scheme with a $2^{nd}$ order Runge-Kutta time stepping (the grid spacing and time increment were $\Delta x=\sigma_0/8$ and $\Delta t=0.025$), respectively \cite{SVC}. 

\subsection{Dynamical response for elastic deformations}

\subsubsection{Compression and torsion of a crystalline cluster}

\begin{figure}
\includegraphics[width=0.33\linewidth]{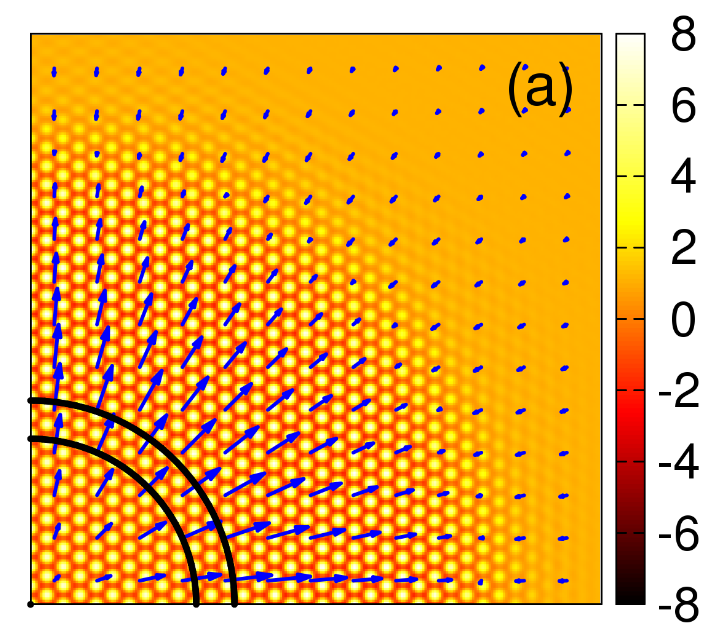}
\includegraphics[width=0.66\linewidth]{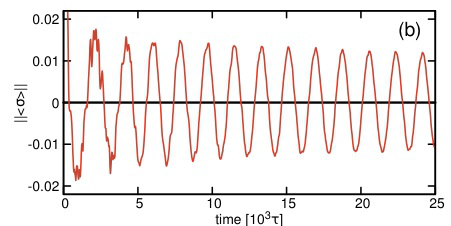}\\
\includegraphics[width=0.33\linewidth]{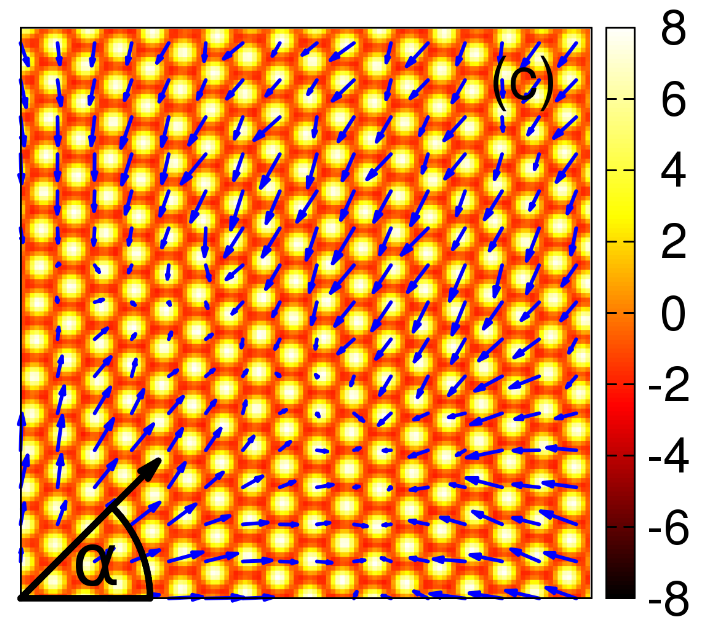}
\includegraphics[width=0.66\linewidth]{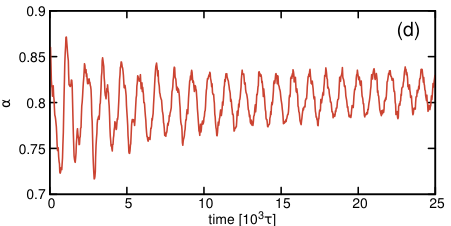}
\caption{Response of a crystalline cluster to deformations. (a) and (c) Typical density distributions (colored maps) and coarse-grained velocity field (arrows) in the case of compression / torsion. (b) Time evolution of the normalized average lattice constant $||\langle \sigma \rangle||=\langle \sigma \rangle/ \sigma_0-1$ [$\langle \sigma \rangle$ measured inside a ring between the two concentric circles in panel (a)] in the case of compression. (d) Time evolution of the angle (relative to the x axis) shown in panel (c) in the case of torsion. For the definition of the vector, see the text.}
\end{figure}

\par First, the elastic response of a deformed 2D crystal was studied. The simulations were performed on a $1024 \times 1024$ grid. The initial condition was a circular crystal grain equilibrated with the surrounding liquid. Next, two types of deformation were applied on the cluster: (i) compression, by radial $h$, and (ii) torsion, by tangential modification of the pattern. In case (i) circular compression waves were observed: the average first neighbor distance (lattice constant) in a thin ring oscillates around its equilibrium value, which is the consequence of the wave dynamics. A similar phenomenon could be observed in case of (ii), where as a response to the torsion the rings have peformed angular oscillations [see figures 1(c) and (d)]. Figure 1(d) displays the time evolution of the angle $\alpha$ shown in figure 1(c) [$\alpha$ is the angle of the vector anchored to an individual particle, i.e. a density maximum]. Our results imply that the inclusion of inertia establishes qualitatively correct dynamics for elastic properties.

\subsubsection{Phonon spectrum}

 \begin{figure}
\includegraphics[width=0.49\linewidth]{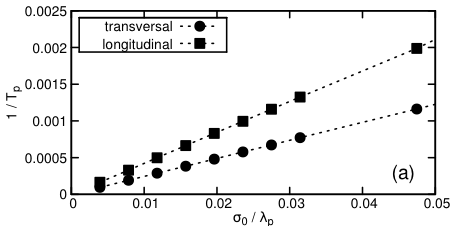}
\includegraphics[width=0.49\linewidth]{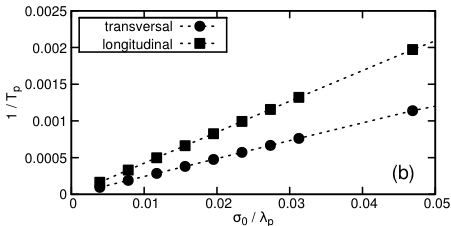}
\caption{Phonon spectrum of a bulk triangular  crystal for the (a) $\mathbf{q}=[q,0]$ and (b) $\mathbf{q}=[0,q]$ directions. Here $T_p$ is the period and $\lambda_p$ the wavelength of the phonon.}
\end{figure}

As a more precise analysis of the dynamic response for elastic deformations, we measured the phonon spectrum of the bulk crystal. In the 2d monatomic triangular  lattice two acoustic phonon branches are present: a longitudinal and a transversal branch. Considering a crystal orientation described by the real-space basis vector set $\{[0,\sigma_0],[\pm(\sqrt{3}/2) \sigma_0,- \sigma_0/2]\}$ the following two dominant phonon modes emerge from the classical (spring-mass harmonic oscillator) model: (i) For $\mathbf{q}=[q,0]$
\begin{eqnarray}
\label{eq:phonon1}\omega_{T}(q) &\propto& \sqrt{ 1-\cos [(\sqrt{3}/2) \sigma_0 q ] } \\
\label{eq:phonon2}\omega_{L}(q) &\propto& \sqrt{ 3 \{ 1-\cos [(\sqrt{3}/2) \sigma_0 q ] \} } \enskip ,
\end{eqnarray}
where $T$ and $L$ denote the transversal ($\mathbf{u}_0 \perp \mathbf{q}$) and longitudinal ($\mathbf{u}_0 \parallel \mathbf{q}$) branches, respectively. (ii) For $\mathbf{q}=[0,q]$, in contrast,
\begin{eqnarray}
\label{eq:phonon3}\omega_T(q) &\propto& \sqrt{3[1-\cos(\sigma_0 q/2)]} \\
\label{eq:phonon4}\omega_L(q) &\propto& \sqrt{3 - \cos(\sigma_0 q/2) - 2\cos(\sigma_0 q)} \enskip .
\end{eqnarray}   
Taylor expanding equations (\ref{eq:phonon1}), (\ref{eq:phonon2}), (\ref{eq:phonon3}) and (\ref{eq:phonon4}) for small wave numbers gives the following linear relationships:
\begin{equation}
\omega_{T,L}^{1,2} = c_{T,L}^{1,2} \cdot q \enskip ,
\end{equation} 
where superscripts $1$ and $2$ refer to the $\mathbf{q}=[q,0]$ and $\mathbf{q}=[0,q]$ directions, respectively, while $c_L^1/c_T^1=c_L^2/c_T^2=\sqrt{3}$ and $c_L^1/c_L^2=c_T^1/c_T^2=1$. 

 As the initial condition for the numerical simulations we used the bulk equilibrium density distribution of the crystal $\rho(\mathbf{r},0)=\rho_S^{eq}(\mathbf{r})$ and a small amplitude momentum perturbation $\mathbf{p}(\mathbf{r},0)=\rho_S^{eq}(\mathbf{r}) \cdot \mathbf{v}_0(\mathbf{r})$, where $\mathbf{v}_0(\mathbf{r})=\mathbf{v_0}\sin[\mathbf{q} \cdot \mathbf{r}]$ (the wavelength $\mathbf{q}$ was commensurable with the system size). After a short transient time the system comes to a coherent phonon state. Measuring the displacement field $\mathbf{u}(\mathbf{r}_i,t)$ by tracking the lattice sites (where $\mathbf{r}_i$ denotes the  equilibrium position of a density peak) gives the time dependent phonon amplitude $\mathbf{u}_0(t)=\mathbf{u}_0 \sin[\omega(\mathbf{q}) t]$, where $\mathbf{u}_0 \parallel \mathbf{v}_0$. The results are summarized in figure 2, where the inverse frequency ($1/T_p$, where $T_p$ is the period of the phonon measured in natural units $\tau$) is plotted against the inverse phonon wavelength $\sigma_0/\lambda_p$ (where $\lambda_p$ is the phonon wavelength). The linear fit $1/T_p = C(\sigma_0/\lambda_p)$ resulted in the following coefficients: 
$C_1^T=0.024555$, $C_1^L=0.042227$, $C_2^T=0.024405$ and $C_2^L=0.042286$. The relative errors of coefficients' ratios compared to the corresponding theoretical ratios of coefficients $c_{L,T}^{1,2}$ are less then $0.7\%$, which means an excellent agreement with the classical theory for phonon wavelengths $\lambda_p > 20 \sigma_0$.

\subsection{Steady front propagation}

\par Next, the propagation of the (10) triangular  planar front has been studied during freezing and melting. The calculations were performed on a $65536\times128$ grid. As expected, steady propagation of the crystal-liquid interface was observed, which follows from the presence of the fluid flow. The front velocity as a function of the relative supersaturation is shown in figure 3(d).
\begin{figure}
\includegraphics[width=0.49\linewidth]{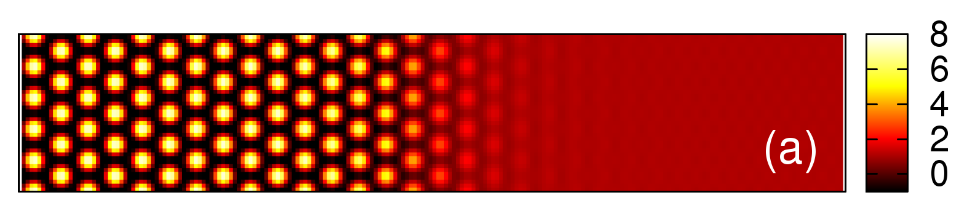}
\includegraphics[width=0.49\linewidth]{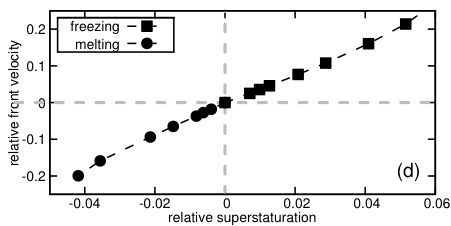}
\includegraphics[width=0.49\linewidth]{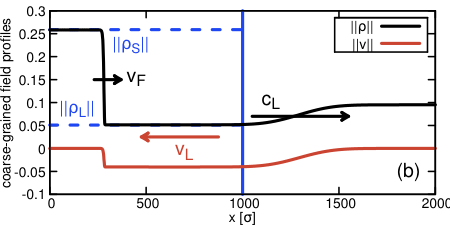}
\includegraphics[width=0.49\linewidth]{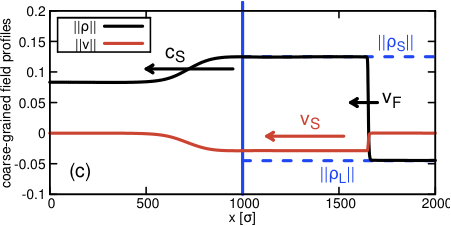}
\caption{Steady crystal growth/melting. (a) Typical density distribution of a crystal-liquid interface (b) typical cross section of the normalized coarse-grained density [$||\rho||=(\hat{\rho}-\rho_L^{eq})/\rho_L^{eq}$] and the $x$ component of the normalized velocity ($||v||=v/c_0$) field in case of freezing, and (c) the same for melting.  (d) Steady front velocity as a function of the relative supersaturation $||\rho_L||$.}
\end{figure}

Figures 3(b) and (c) show typical cross sections of the coarse-grained density during crystallization and melting, respectively. In both cases a density wave travels ahead of the crystal-liquid front at close to the speed of sound in the liquid and the crystal. Behind the density wave a steady convective mass flux evolves (denoted by $v_L$ and $v_S$ in the figures), establishing the required mass flux for freezing, or the removal of the excess material in the case of melting. The velocity of the crystal-liquid front satisfies mass continuity in both cases: $(\hat{\rho}_S-\hat{\rho}_L) \cdot v_F=\hat{\rho}_{L} \cdot v_{L}$ for freezing, and $(\hat{\rho}_S-\hat{\rho}_L) \cdot v_F=\hat{\rho}_{S} \cdot v_{S}$  for melting.

 It is worth noting that the analytical results of Refs. \cite{GE,ARTK} for the steady front velocity refer to the unstable liquid that exists beyond the linear stability line, which lies far beyond the supersaturation regime covered by our simulations. The assumptions made in deriving the analytical expression are not valid in the latter regime. Unfortunately, we were unable to get a steady state planar front for higher supersaturations due to a stress induced destabilization of the growth front at large supersaturations, where the dynamically forming lattice constant differs from the equilibrium lattice constant, which in turn was used in setting the width of the long simulation box. Evidently, carefully designed simulations may overcome this difficulty, however, such investigations are left for a future publication. We wish to stress, furthermore, that the front velocities deduced in Refs. \cite{GE,ARTK} apply best probably for describing the 'fast' diffusionless mode of solidification observed close to the instability line \cite{TTG,TGTDP}, where indeed the density change upon solidification is negligible \cite{TTG,TGTDP}.  

\subsection{Capillary wave analysis}

The capillary wave spectrum of the fluctuating 2D crystal-liquid equilibrium interface has been analyzed in the presence of fluctuations. In order to exclude subatomic level noise components, the noise field is colored, namely, $S_{ij}(\mathbf{k})\equiv 0$ for $|\mathbf{k}|>k_{max}$, where $k_{max}<k_0/2$. Accordingly, the numerical covariance tensor  of the noise matrix  corresponding to equation (\ref{eq:eq7}) has the prefactor $\Gamma/(\Lambda^3\Delta t)$ \cite{GSR}, where $\Lambda=(k_0/k_{max})(\sigma_0/2)$ is the correlation length of the noise consistently with the crystal structure, whereas the factor $\Gamma=(2\pi/\Delta x)^3/[(4\pi/3)k_{max}^3]$ establishes the constant total power of the noise independently from $k_{max}$ (The power spectral density of a Gaussian white noise is constant, therefore, the total power is simply proportional to the number of discretized wave-vectors.). The simulations were performed on a 1024$\times$4096 grid, with an equilibrium liquid-crystal-liquid slab as the  initial condition. Figure 4(a) shows a snapshot of the fluctuating $(11)$ triangular  crystal-liquid interface (only part of the simulation box is presented here). The white curve shows the interface position $h(x,t)$ calculated from the lattice site positions (locations of local maxima of the density distribution above a threshold value $\Pi$) In our calculations $\Pi=4$ was chosen. On the basis of the capillary wave theory \cite{BL} the interface stiffness can be measured by investigating the long wavelength behavior of the capillary wave spectrum: $\langle |h(q)|^2 \rangle = (k_B T)/(q^2 \tilde{\gamma} L)$, where $\langle . \rangle $ denotes time averaging. The interface stiffness reads as: $\tilde{\gamma}=\gamma(\theta)+\gamma''(\theta)$, where $\gamma(\theta)$ is the line free energy (2 dimensional interfacial free energy) as a function of the orientation of the crystal face, and $\gamma''(\theta)$ its second derivative with respect to the orientation. $L$ denotes size of the box in $x$ direction. Figure 4(b) shows the capillary wave spectrum of the $(11)$ interface, which is in excellent qualitative agreement with the result of molecular dynamics simulations \cite{RH}. From the long wavelength fitting $\log[\langle |h(q)|^2 \rangle]= -2q + \log(\tilde{\gamma}L)$ the stiffness $\tilde{\gamma}$ can be determined as a function of the crystal plane orientation. For weak anisotropies [the equilibrium crystalline cluster is almost circular, see figure 1(a)], the anisotropic line free energy of a triangular crystal can be approximated as $\gamma(\theta)=\gamma_0[1+\epsilon \cos(6\theta)]$.The anisotropy parameter can be evaluated from the simulations as $\epsilon=|(\gamma_{max}-\gamma_{min})/(\gamma_{max}+\gamma_{min})|$. Unfortunately, the anisotropy of the present system is comparable to the error of fitting $\log(L\tilde{\gamma})$, therefore, only a rough estimate of the anisotropy parameter can be obtained: Assuming $\tilde{\gamma}=\gamma_0[1-35\epsilon\cos(6\theta)]$ the anisotropy parameter is approximated as $\epsilon \approx |(\tilde{\gamma}_{max}-\tilde{\gamma}_{min})/(\tilde{\gamma}_{max}+\tilde{\gamma}_{min})|/35$. Measuring the stiffness for 10 differently oriented crystal faces resulted in $\epsilon \lesssim 0.002$, which is somewhat smaller than in the noiseless case, where a direct evaluation of the line free energy \cite{TTPTG} yields $\approx 0.0026$ \cite{T}. It is worth noting that at the same reduced temperature the anisotropy of the bcc-liquid system is larger by an order of magnitude \cite{PTPG}, suggesting an anisotropy of $1-2\%$, in agreement with molecular dynamics results for metallic systems \cite{Asta}.\\consistent

\begin{figure}
\includegraphics[width=0.49\linewidth]{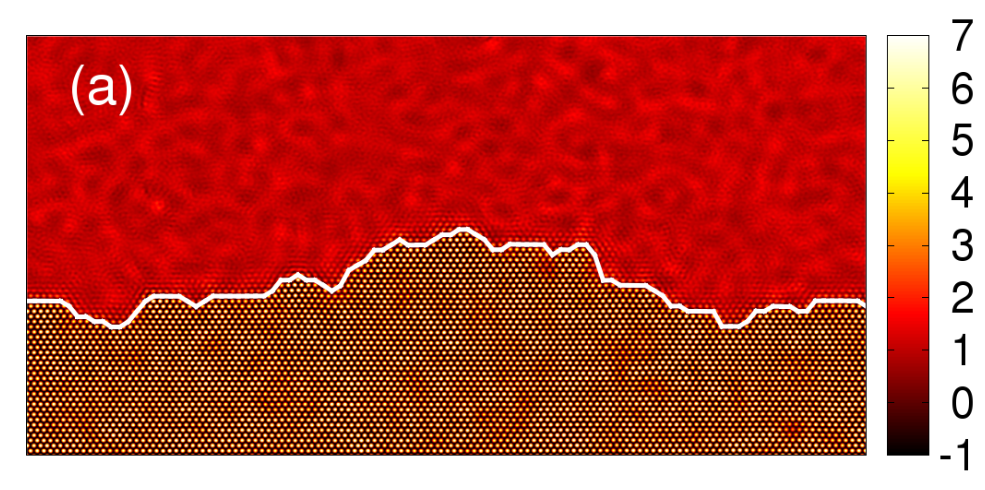}
\includegraphics[width=0.49\linewidth]{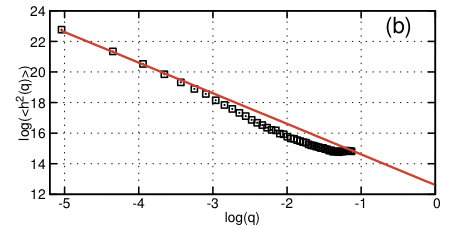}
\caption{Capillary wave analysis of the $(11)$ triangular  crystal face. (a) Snapshot of the fluctuating crystal-liquid interface. The interface position [$h(x)$] is denoted by the solid white line. (b) Log-log plot of the capillary wave spectrum $\langle h^2(q) \rangle$ (black squares) and the long wavelength linear fit (solid red line).}
\end{figure}

\subsection{Growth vs. shear- and bulk viscosity}

We address here the dependence of the crystal growth rate on the shear viscosity ($\mu_S$) and bulk viscosity ($\mu_B$). While theoretical considerations and molecular dynamics simulations suggest that the front velocity scales with $\mu_S^{-1}$ in the vicinity of the melting point, a smaller negative exponent is reported at large undercoolings \cite{EHY}. Shear viscosity can be measured with relative ease in a broad temperature range via combining various methods, and is well known for a range of metallic systems \cite{BG,CS}. Much less is known about the bulk viscosity, mainly due to the complexity and the associated uncertainty of the measurement techniques \cite{LPR}. The handful of relevant experimental data for metallic systems vary between $\mu_B/\mu_S=1/2$ and $8$ \cite{FGL}. Although this is a quite broad range, the effect on the front velocity is expected to be small in systems with small density gaps (e.g., when the divergence of the velocity field is moderate). 

First, we investigate the dependence of front velocity on shear viscosity. The results are summarized in figure 5(a). For constant relative supersaturation ($||\rho_L|| \approx$ 0.01) we recover the expected $v_F \propto \mu_S^{-1}$ relationship for $\mu_B/\mu_S = 2$.

Next, we study the effect of bulk viscosity on front propagation in the range of $\mu_B/\mu_B^0=1/16$ to $8$ at constant $\mu_S=\mu_S^0$. As indicated by figure 7(b), the front velocity is a "weak" function of the bulk viscosity: $(v_F-v_F^0)/v_F^0 \propto \log(\mu_B/\mu_{B,0})$, where $\mu_B^0=2\mu_S^0$ is the reference bulk viscosity and $v_F^0$ the corresponding front velocity. Remarkably, a factor of $16$ in the bulk viscosity results in only a factor less than $2.5$ in the front velocity. In addition, in real metallic systems the dimensionless compressibility of the liquid is smaller by a factor of 4 than the $C_0$ used here, therefore, the respective physical density gap is less than $3\%$ \cite{JAEA}, suggesting an even smaller effect of the bulk viscosity on the steady-state front velocity.

\begin{figure}
\includegraphics[width=0.49\linewidth]{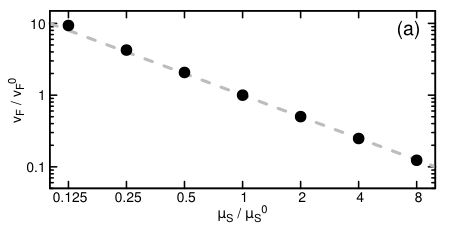}
\includegraphics[width=0.49\linewidth]{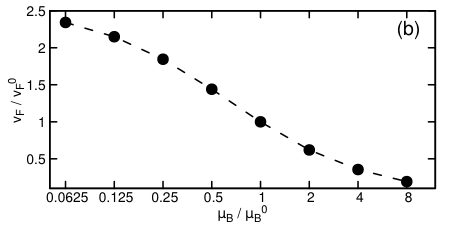}
\caption{Effect of (a) shear viscosity ($\mu_S$)  and (b) bulk viscosity ($\mu_B$) on the front velocity ($v_F$) in the case of steady front propagation. A relative supersaturation of $\approx 0.01$ applies.}
\end{figure}

\subsection{The effect of the coarse-graining filter}

\begin{figure}
\includegraphics[width=0.31\linewidth]{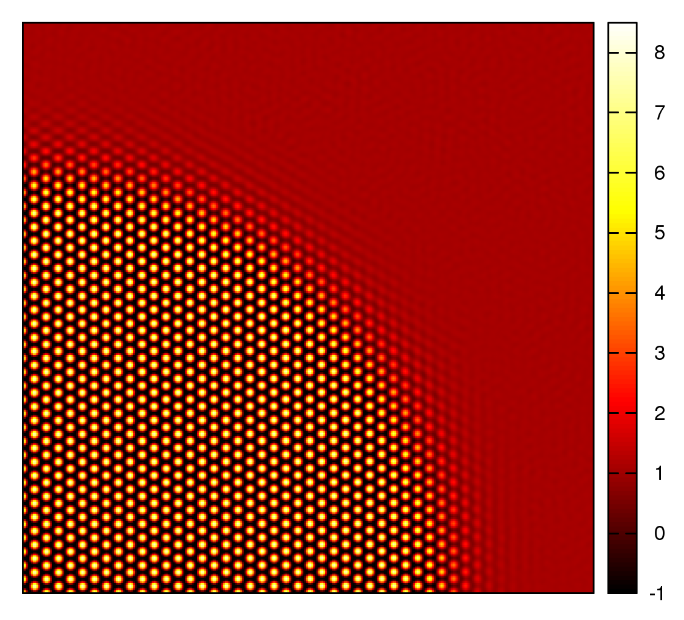}
\includegraphics[width=0.33\linewidth]{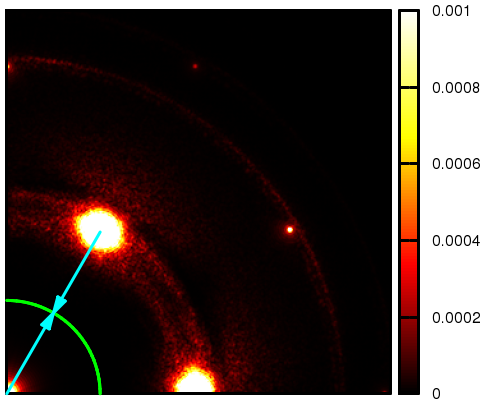}
\includegraphics[width=0.32\linewidth]{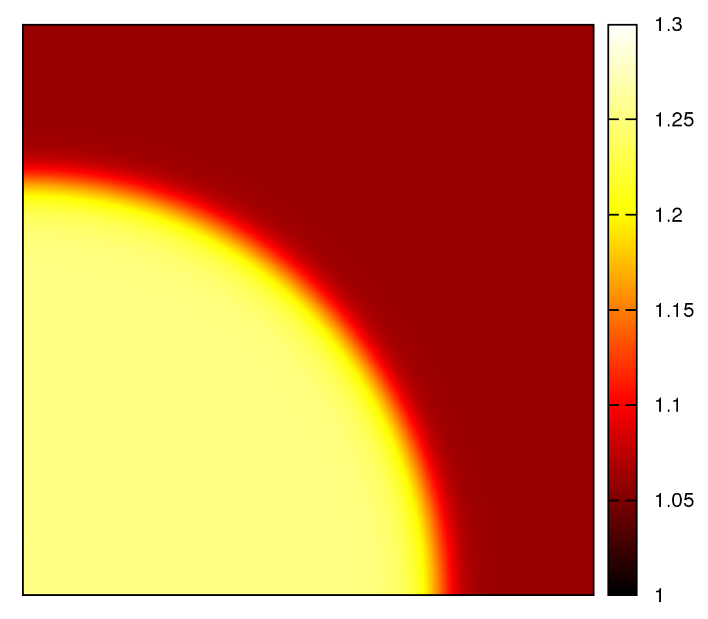}
\caption{Coarse-graining of the density field. (a) Original density distribution of an equilibrium crystalline cluster (top-right quarter of the computational domain). (b)  Spectrum [magnitude of $\tilde{\rho}(q)$] of the full density distribution. The coarse-grained density is generated by applying various lowpass filters with cutoff length $t=1/2$ (denoted by the green arc). (c) Coarse-grained density field corresponding to panel 'a'.}
\end{figure}

Finally, we explore how the results depend on the choice of the coarse-graining filter. Evidently, the coarse-grained density (and momentum) must be flat in the bulk phases. The coarse-grained fields are generated by using the following convolution \cite{FIR}:

\begin{equation}
\hat{\phi}(\mathbf{r},t) := \int d\mathbf{r}' \left\{ K(\mathbf{r}-\mathbf{r}')\phi(\mathbf{r}',t) \right\} \enskip ,
\end{equation}
where $\phi$ can be either the density or the momentum fields, and $K(\mathbf{r})$ is the kernel of the filter. In our calculations we used various lowpass filters. The spectral kernel of a generalized quasi-ideal lowpass filter reads as
\begin{equation}
\label{eq:filter_lowpass}
K(q)=\{1-\tanh[(q-q_0)/\xi]\}/2 \enskip ,
\end{equation}
where $\xi$ is a small parameter. Note that for $\xi\to0$ this filter recovers the ideal lowpass filter: $K(q)=\theta(q-q_0)$, where $\theta(x)$ is the Heaviside function. Another possible choice can be the Gaussian filter, for which
\begin{equation}
\label{eq:filter_gauss}
K(q)=\exp[-(q s)^2/2] \enskip ,
\end{equation}
where the real-space filtering length, $s$ (i.e. the standard deviation), emerges from symmetry considerations as follows: the density can be approximated as 
\begin{equation}
\label{eq:rhoapprox}
\rho(\mathbf{r}) \approx \rho_0+f(\mathbf{r}) [\Delta\rho+ \sum_i A_i \exp(\imath \mathbf{\Gamma}_i \mathbf{r} ) ] \enskip ,
\end{equation}
 where we have a lattice periodic function (plus a density gap $\Delta\rho$) modulated by the slowly varying envelope function $f(\mathbf{r})$ [as illustrated in figure 6(a)], while the corresponding coarse-grained density reads as $\hat{\rho}(\mathbf{r})=\rho_0+f(\mathbf{r})\Delta\rho$ [see figure 6(c)]. The slow variation of $f(\mathbf{r})$ means that its spectrum decays relatively fast around $q=0$. Because of the discrete sum in Eq. (\ref{eq:rhoapprox}) in the spectrum of the full density [$\tilde{\rho}(\mathbf{q})$] the spectrum of the envelope function [$\tilde{f}(\mathbf{q})$] is simply repeated around every reciprocal lattice vector [see figure 6(b)]:
 \begin{equation}
 \tilde{\rho}(\mathbf{q}) = \rho_0 \cdot \delta(0)+\Delta\rho \tilde{f}(\mathbf{q})+ \sum_i A_i \cdot  \tilde{f}(\mathbf{q}-\mathbf{\Gamma}_i) \enskip ,
 \end{equation}
 where $\delta(0)$ is the delta function. Choosing one of the first reciprocal lattice vectors ($\mathbf{q}_0$) and taking the cross section of the spectrum at $\mathbf{q}=t \cdot \mathbf{q}_0$ one can define the functions $g(t):=|\tilde{f}(t \cdot \mathbf{q}_0)|$ and $h(t):=|\tilde{\rho}(t \cdot \mathbf{q}_0)|$. If $g(t)$ is a fast decaying function, then $h(t)$ decays around $t=0$ and $t=1$ as $\Delta \rho \cdot f(t)$ and $A_0 \cdot f(1-t)$, respectively [as indicated by the vectors in figure 6(b)]. Therefore, the expected optimum spectral cutoff distance for the filters is at $t_0=\Delta \rho/A_0$. In our case the density gap is $\Delta \rho \approx 0.2$, while the amplitude is $A_0 \approx 1$, therefore, $t_0\approx 0.2$.
 
 Although the approximation described above might seem to be too simple and rough, it has been verified numerically for the Gaussian filter. We have investigated both $h(t)$ and the coarse-grained density in the crystal as a function of the real space filtering length $s$ (standard deviation)  for an equilibrium crystalline cluster. Figure 7.a shows h(t) for various filtering lengths $s/\sigma_0$. It can be clearly seen that the spectrum is practically independent from $s$. The best choice for $s$ is then determined by the flatness of the coarse-grained density in the bulk crystal and the behavior of the solid-liquid interface [see figure 7(b)]: If the filtering length is too large, the interface broadens (e.g., for $s/\sigma_0=2.5,3.0,$ and $3.5$), since the filter smoothes the density. In contrast, if $s$ is too small (e.g., for $s/\sigma_0=1.2,1.4,1.6,$ and $1.8$), significant periodic components appear in the bulk crystal. As indicated by figure 7(b), an optimum choice for the real-space filtering length can be $s=2 \sigma_0$, for which the spectral kernel of the Gaussian filter cuts off for $t=q/q_0 \gtrsim 0.2$ [see figure 7(c)]. 

\begin{figure}
\includegraphics[width=0.49\linewidth]{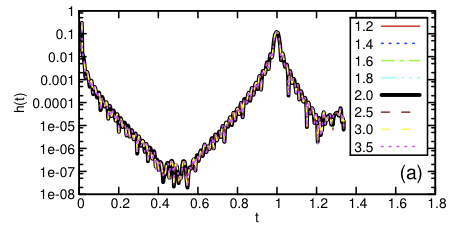}
\includegraphics[width=0.49\linewidth]{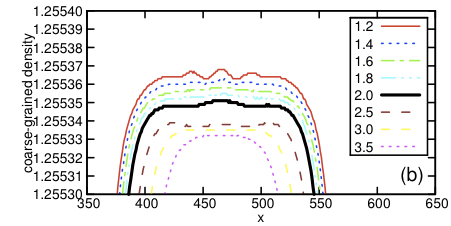}
\includegraphics[width=0.49\linewidth]{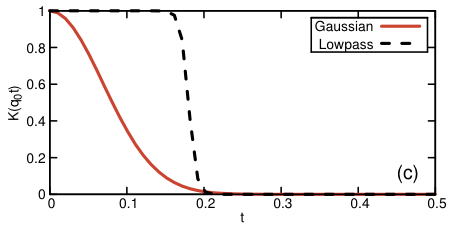}
\includegraphics[width=0.49\linewidth]{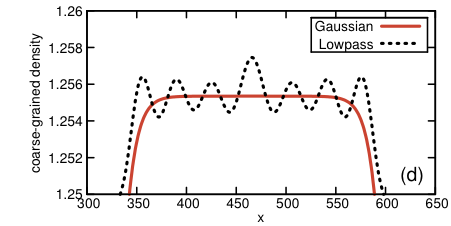}
\caption{Effect of the coarse-graining filters. (a) Cross section of the spectrum of the full density for various real-space filtering lengths (standard deviation in real space) in case of the Gaussian filter: $s/\sigma_0=1.2,1.4,1.6,1.8,2.0,2.5,3.0,$ and $3.5$ [For the definition of $h(t)$ see the text.]. (b) Cross section of the coarse-grained density for the same filtering lengths. (c) Spectral kernels of different coarse-graining filters: the Gaussian filter for $s=2 \sigma_0$ and the lowpass filter for $q_0=0.18$ and $\delta=0.01$. (d) Comparison of the coarse-grained density for the filters shown by panel c. The density profiles were obtained from the cross section of the equilibrium cluster shown in figure 6. Note the difference between the 'y' axis scales of panel b and d, which indicates a huge error of the lowpass filter compared to all the Gaussians.}
\end{figure}

As an alternative to the Gaussian filter, one may wish to use a generalized quasi-ideal lowpass filter defined by equation (\ref{eq:filter_lowpass}). The problem with that filter has already been illustrated in figure 7(a): The spectrum decays exponentially around $t=0$ and $t=1$, and there is no "empty" frequency range of finite size between them. Unfortunately, the quasi ideal lowpass filters are not smoothing filters, rather they are developed to separate two distinct (non-overlapping) frequency regimes. Therefore, the expected quality of the coarse-grained density is much worse in case of these filters, which also has been verified numerically [see figure 7(d)]. Consequently, in our simulations we used a Gaussian filter with a filtering length $s=2 \sigma_0$, as indicated above.

\section{Summary}

We have presented a nonlinear hydrodynamic density functional theory of crystallization. Our work couples the Classical Density Functional Theory to the Navier-Stokes equation directly via a phenomenological coarse-graining technique that prevents velocity singularities in the interatomic space. The dynamical equations are fundamental and supplemented with momentum fluctuations emerging from the fluctuation-dissipation theorem. The main virtue of the present model is that the elastic of the crystal and the liquid is incorporated consistently. This is demonstrated by illustrative simulations including dynamic response to elastic deformations, phonon spectrum and steady crystal growth/melting. In addition, the capillary wave spectrum and the anisotropy of the interfacial free energy are also in a good qualitative agreement with the results of atomistic simulations. Our work opens up the way to address a range of microscopic phenomena, such as homogeneous and heterogeneous nucleation, morphology evolution, kinetic roughening, fragmentation, etc. in forced convection.

\section*{Acknowledgement}
This work has been supported by the EU FP7 Collaborative Project “EXOMET” (contract no. NMP-LA-2012-280421, co-funded by ESA) and by the ESA MAP/PECS project “GRADECET” (Contract No. 4000104330/11/NL/KML). G. Tegze is a grantee of the J\'anos Bolyai Scholarship of the Hungarian Academy of Sciences.


\section*{References}
\begin{thebibliography}{abbrv}
\bibitem{E} Evans R 1979 {\it Adv. Phys.} {\bf 28} 143
\bibitem{O} Oxtoby D W 1991 in {\it Liquids, Freezing and the Glass Transition} (North Holland, Amsterdam, 1991), vol. 1, p. 145.
\bibitem{Si} Singh Y 1992 {\it Phys. Rep.} {\bf 207} 351
\bibitem{RY} Ramakrishnan T W and Yussouff M 1979 {\it Phys. Rev. A} {\bf 19} 2755
\bibitem{WDA} Curtin W A and Ashcroft N W 1985 {\it Phys. Rev. A} {\bf 32} 2909
\bibitem{MWDA} Denton A R and Aschcroft N W 1989 {\it Phys. Rev. A} {\bf 39} 4701
\bibitem{FMT} Rosenfeld Y 1989 {\it Phys. Rev. Lett.} {\bf 63} 980
\bibitem{PFC0} Elder K R, Katakowski M, Haataja M and Grant M 2002 {\it Phys. Rev. Lett.} {\bf 88} 245101
\bibitem{PFC1} Elder K R, Provatas N, Berry J, Stefanovic P and Grant M 2007 {\it Phys. Rev. B} {\bf 75} 064107
\bibitem{AP} Emmerich H, L\"owen H, Wittkowski R, Gruhn T, T\'oth G I, Tegze G and Gr\'an\'asy L 2012 {\it Adv. Phys.} {\bf 61} 665
\bibitem{HGL} Haataja M, Gr\'an\'asy L and L\"owen H 2010 {\it J. Phys.: Condens. Matter} {\bf 22} 360301
\bibitem{MT} Marconi U M B and Tarazona P 1999 {\it J. Chem. Phys.} {\bf 110} 8032
\bibitem{AE} Archer A J and Evans R 2004 {\it J. Chem. Phys.} {\bf 121} 4246
\bibitem{AR} Archer A J and Rauscher M 2004 {\it J. Phys. A: Math. Gen.} {\bf 37} 9325
\bibitem{TLL} van Teeffelen S, Likos C N and L\"owen H 2008 {\it Phys. Rev. Lett.} {\bf 100} 108302
\bibitem{A06} Archer A J 2006 {\it J. Phys.: Condens. Matter} {\bf 18} 5617
\bibitem{A09} Archer A J 2009 {\it J. Chem. Phys.} {\bf130} 014509
\bibitem{TBVL} van Teeffelen S, Backofen R, Voigt A and L\"owen H 2009 {\it Phys. Rev. E} {\bf 79} 051404
\bibitem{SHP} Stefanovic P, Haataja M and Provatas N 2006 {\it Phys. Rev. Lett.} {\bf 96} 225504
\bibitem{MG} Majaniemi S and Grant M 2007 {\it Phys. Rev. B} {\bf 75} 054301
\bibitem{MNG} Majaniemi S, Nonomura M and Grant M 2008 {\it Eur. Phys. J. B} {\bf 66} 329
\bibitem{PV} Preatorius S and Voigt A 2011 {\it Macromol. Theory Simul.} {\bf 20} 541
\bibitem{ALV11} Aland S, Lowengrub J and Voigt A 2011 {\it Phys. Fluids} {\bf 23} 062103
\bibitem{ALV12} Aland S, Lowengrub J and Voigt A 2012 {\it Phys. Rev. E} {\bf 86} 046321
\bibitem{SVC} Shang B Z, Voulgarakis N K and Chu J.-W., 2011 {\it J. Chem. Phys.} {\bf 135} 044111
\bibitem{LL} Landau L D and Lifshitz E M, 1959 {\it Fluid Mechanics} (Pergamon, New York)
\bibitem{NY} Nakamura T and Yoshimori A 2009 {\it J. Phys. A: Math. and Theor.} {\bf 42} 065001
\bibitem{Sa} Salmon R 1988 {\it Annu. Rev. Fluid Mech.} {\bf 20} 225
\bibitem{LiLi} Liu C and Li Z 2011 {\it AIP Adv.} {\bf 1} 032108 
\bibitem{EG} Elder K R and Grant M 2004 {\it Phys. Rev. E} {\bf 70} 051605
\bibitem{TGTPJAP} Tegze G, Gr\'an\'asy L, T\'oth G I, Podmaniczky F, Jaatinen A, Ala-Nissila T and Pusztai T 2009 {\it Phys. Rev. Lett.} {\bf 103} 035702
\bibitem{TTG} Tegze G, T\'oth G I and Gr\'an\'asy L 2011 {\it Phys. Rev. Lett.} {\bf 106} 195502
\bibitem{TPTTG} T\'oth G I, Pusztai T, Tegze G, T\'oth G and Gr\'an\'asy L 2011 {\it Phys. Rev. Lett.} {\bf 107} 175702
\bibitem{TTPG} T\'oth G I, Tegze G, Pusztai T and Gr\'an\'asy L 2012 {\it Phys. Rev. Lett.} {\bf 108} 025502
\bibitem{BG} Berry J and Grant M 2011 {\it Phys. Rev. Lett.} {\bf 106} 175702
\bibitem{HHP} Humadi H, Hoyt J J and Provatas N 2013 {\it Phys. Rev. E} {\bf 87} 022404
\bibitem{JAEA} Jaatinen A, Achim C V, Elder K R and Ala-Nissila T 2009 {\it Phys. Rev. E} {\bf 80} 031602
\bibitem{Ka} Kaptay G 2005 {\it Z. Metallkd.} {\bf 96} 42
\bibitem{GSR} Garcia-Ojalvo J, Sancho J M and Ramirez-Piscina L 1992 {\it Phys. Rev. A} {\bf 46} 4670
\bibitem{GE} Galenko P K and Elder K R 2011, {\it Phys. Rev. E} {\bf 83} 064113
\bibitem{ARTK} Archer A J, Robbins M J, Thiele U and Knobloch E 2012, {\it Phys. Rev. E} {\bf 86} 031603
\bibitem{TGTDP} Tegze G, Gr\'an\'asy L, T\'oth G I, Douglas J F and Pusztai T 2011, {\it Soft Matter} {\bf 7} 1789
\bibitem{BL} Buff F P, Lovett R A and Stillinger F H 1965, {\it Phys. Rev. Lett.} {\bf 15} 621
\bibitem{RH} Rozas R E and Horbach J 2011 {\it Europhys. Lett.} {\bf 93} 26006
\bibitem{TTPTG} T\'oth G I, Tegze G, Pusztai T, T\'oth G and Gr\'an\'asy L 2010 {\it J. Phys.: Condens. Matter} {\bf 22} 364101
\bibitem{T} T\'oth G I, to be published
\bibitem{PTPG} Podmaniczky F, T\'oth G I, Pusztai T and Gr\'an\'asy L 2013 {\it J. Cryst. Growth}, published electronically. http://dx.doi.org/10.1016/j.jcrysgro.2013.01.036
\bibitem{Asta} Asta M, Hoyt J J and Karma A 2003 {\it Mater. Sc. Eng. R} {\bf 41} 121
\bibitem{EHY} See for example, Ediger M D, Harrowell P and Yu L 2008 {\it J. Chem. Phys.} {\bf 128} 034709, and references therein.
\bibitem{BG} Battezzati L and Greer A L 1989 {\it Acta Metall.} {\bf 37} 1791
\bibitem{CS} Chhabra R P and Sheth D K 1990 {\it Z. Metallkde.} {\bf 81} 264 
\bibitem{LPR} Liebermann R N 1949, {\it Phys. Rev.} {\bf 75} 1415
\bibitem{FGL} Flinn J M, Gupta P K and Litovitz T A 1974 {\it J. Chem. Phys.}  {\bf 60} 4390
\bibitem{FIR} Press W H, Teukolsky S A, Vetterling W T, and Flannery B P 1992, Numerical Recepes in C, 2nd ed. Cambridge University Press, Cambridge
\end{thebibliography}
\end{document}